\documentclass[a4paper,10.5pt]{article}
\usepackage[dvips]{graphicx}

\begin{document}

\begin{flushleft}

{\Large{\bf ASCA and XMM-Newton Observations of A2029}} \\

\bigskip
\bigskip
\smallskip

S. Miyoshi$^{\rm a}$, N. Tanaka$^{\rm a}$, M. Yoshimura$^{\rm a}$, K. Yamashita$^{\rm b}$, A. Furuzawa$^{\rm b}$, T. Futamura$^{\rm b}$, and M. Hudaverdi$^{\rm b}$

\bigskip
\smallskip

$^{\rm a}Department \ of \ Physics, \ Kyoto \ Sangyo \ University, Kita-ku, \ Kyoto \ 603-8555, \ Japan$ \par
$^{\rm b}Department \ of \ Physics, \ Nagoya \ University, \ Chikusa-ku, \ Nagoya \ 464-8602, \ Japan$ \par
\end{flushleft}
\noindent{\bf Abstract}\par
\bigskip
\noindent The X-ray data of A2029 obtained with XMM-Newton show no evidence of an embedded AGN in the central region of this cluster, which was suggested from the analysis of restored ASCA image data, although some hot spots are seen within or around the central cD galaxy. The absence of AGN at the cluster center is consistentent with the result of Chandra observations. Radial profiles of the iron abundance and the 2D (surface) temperature obtained from the XMM-Newton data are in good agreement with the Chandra data as a whole. \\

\bigskip

\noindent{\bf 1. Introduction} \\

\smallskip

A2029 is a nearby ($z=0.0767$) rich cluster of galaxies. This cluster has been observed with ROSAT (January 12, 1992), ASCA (February 19, 1994), BeppoSAX (February 4--5, 1998), Chandra (April 12, 2000) and XMM-Newton (August 25, 2002), and so far many papers have been published for these observational results (e.g. Sarazin et al. 1998; Molendi \& De Grandi 1999; Lewis et al. 2002). An interesting result was obtained from the ASCA data. That is, the map of the hardness ratio of this cluster obtained from the restored soft and hard images suggests the existence of an embedded AGN in the central region of this cluster (Tanioka and Miyoshi 2001). While Chandra found no point source in the central region of this cluster (Lewis et al. 2002), there still exists a possibility that XMM-Newton can detect the point source because the X-ray spectrum of the presumed AGN was estimated to be very hard from the ASCA data. Here we report the results of data analysis using the archival data of XMM-Newton together with ASCA results.\par
Throughout this paper, we assume in general a cosmology of $H_0=50\ {\rm km/s/Mpc}, \ \Omega_{\rm m}=1.0, \ \Omega_{\Lambda}=0$. The luminosity distance of this cluster is 468 Mpc (1.96 kpc/arcsec). \\

\bigskip

\noindent{\bf 2. Result of ASCA Observations}\\

\smallskip

Here we summarize the result of our analysis of ASCA data more minutely than before (Tanioka and Miyoshi 2001). The total exposure time of ASCA observations of A2029 was $\sim 4\times 10^4$ s and the detected photon number was $\sim 7.6\times 10^4$. At first we restored the ASCA soft (0.7--1.5 keV) and hard (3--10 keV) X-ray images of A2029 using the Richardson-Lucy (R-L) method.
Since the point spread function is energy-dependent, we adopted a rather narrow bandpass (3--10 keV) to restore the ASCA hard X-ray map.
 The radial profiles of the restored surface brightness of soft and hard X-rays are given in Figs. 1a and 1b. The surface brightness distribution of soft X-rays obtained from the ROSAT data (Sarazin et al. 1998) is quite similar to Fig. 1a and the parameter values for the beta model fit are in good agreement with each other.
 It means that the R-L method we used for the ASCA data worked successfully at least for the restoration of the soft X-ray image. Fig. 2 is the radial profile of the hardness ratio (3--10 keV/0.7--1.5 keV) obtained from the restored image data. The intracluster gas temperature determined from the hardness ratio is consistent with Sarazin et al.'s (1998) result as a whole, except for the central region occupied by the cD galaxy.
 At the center ($r<1'$), the hardness ratio becomes very high and the corresponding temperature of thermal bremsstrahlung exceeds 15 keV. Such a strong hard component suggests the existence of an AGN. Indeed the energy spectrum is well fitted to the sum of a thermal bremsstrahlung of $kT=7.7$ keV (the gas temperature in outer regions) and a power law of $\Gamma =1.5\pm 0.1$ with large absorption ($N_{\rm H}\sim 10^{22}$). The intrinsic X-ray luminosity of the power-law component is estimated to be $L_{\rm X}(2-10 {\rm keV})\sim 5\times 10^{43}{\rm erg/sec}$, which is typical for AGNs.
Such a large absorbing column in the central region of A2029 is not reported by others (Allen \& Fabian 1997; Lewis et al. 2002; Clarke et al. 2004). It is still possible, however, that the absorbing matter is concentrated at around the AGN and a considerable fraction of the hard X-rays from the central region would originate from the AGN embedded in the absorbing matter.\par
The total mass of A2029 is estimated to be $1.17\pm 0.03\times 10^{15}M_{\odot}$ and the mass of the intracluster gas is $1.57\pm 0.03\times 10^{14}M_{\odot}$. The gas fraction of A2029, $0.134\pm 0.04$, is very close to the mean value for clusters of galaxies observed so far (Grego et al. 2001). The baryon mass fraction, $0.209\pm 0.06$, where the total stellar mass $M_{\star}=2.2\times 10^{13}M_{\odot}$ has been estimated from the optical total luminosity of A2029 (Dressler 1979) assuming the stellar mass-to-luminosity ratio of $2M_{\odot}/L_{\odot}$, is nearly equal to that of a flat, accelerated-expansion model of the universe with $\Omega_{\rm M}=0.28$ and $\Omega_{\Lambda}=0.72$. But, it is \,$\sim3$ times larger than that of a flat universe with $\Omega_{\Lambda}=0$.

\bigskip

\noindent{\bf 3. Result of XMM-Newton Observations}\\

\smallskip

The total exposure time of XMM-Newton observations of this cluster is 17,846 sec and the detected total photon numbers are 663,390 (MOS1), 670,300 (MOS2), and 2,351,400 (PN), and about 1/5 of them are suitable for data analysis. Fig. 3 is the XMM-Newton X-ray image of A2029 overlaid by the contour map (logarithmic scale). The position of the cluster center was determined at $\alpha = 15^{\rm h}10^{\rm m}56.^{\rm s}04, \ \delta = 05^{\circ}44'42.''68$ from the X-ray surface brightness distribution. Fig. 4 shows the radial profile of the X-ray surface brightness fitted by the isothermal $\beta$ model. The map of the hardness ratio (3--10 keV/0.7--1.5 keV) for the very central region ($r<17''$) indicates no evidence of point source, although some hot spots are found within or around the cD galaxy (Fig. 5).  The hardness ratio obtained from the same energy bands as in the case of ASCA data mentioned above increases (not decreases) with radius (Fig. 6). These facts show that the temperature increases with radius in the central region. This is clearly inconsistent with the profile we obtained from the ASCA data. Since the angular resolution of XMM-Newton is much better than ASCA, we have to prefer the result obtained from XMM-Newton data to that from ASCA data. It is considered here that our restoration of ASCA images using the R-L method was incomplete for hard X-rays and too many hard X-ray photons were gathered at the centeral region of the restored image. \par
Fig. 7 is the map of the hardness ratio over the whole cluster obtained from the XMM-Newton data with the vignetting correction. The region of $1.5'\le r \le 2.5'$ is very patchy, i.e. high and low temperature regions are jumbled together there in contrast with other regions. The patchiness might be simply seeing noise in the outer region, but it is considered to be significant in the inner region since high values of the hardness ratio there exceed 2.5 $\sigma$ in general. This suggests that the state of intracluster gas is not simple but considerably complicated. This region would be a transition zone between the outer region dominated by matter infalling from the outer space and the inner region in thermal equilibrium, and the patchy high-hardnass-ratio fragments would be in some violent phenomena caused by the infalling matter. \par
Fig. 8 is the 2D (surface) temperature profile obtained from the analysis of energy spectrum including the vignetting correction. Crosses indicate the gas temperatures obtained from the fitting for the whole energy range (0.3--10 keV), and dashed crosses are those for the high energy (2--10 keV) band. The temperature at the center is $4.35\pm 0.35$ keV and increases with radius. All the temperatures indicated by dashed crosses are higher than those indicated by crosses. In addition, the degree of spectral fitting becomes worse at outer regions, when we try to fit the spectrum of 0.3-10 keV band by the thermal bremsstrahlung only. These facts shows that some additional component(s) should be added to the thermal bremsstrahlung component in order to get a good fit to the enegy spectra of outer regions. At the central region, however, the energy spectra are well fitted to the thermal bremsstrahlung by single temperature fit (reduced $\chi ^2 = 1.23$) as well as by multi temperature fit (reduced $\chi ^2 = 1.13$). Fig. 9a shows the energy spectrum of the central region ($r\le 17'$) fitted by the sum of three thermal bremsstrahlung components of 3.1 keV, 4.8 keV and 8.2 keV. These temperatures are taken from the 3D temperature distribution obtained from the Chandra data (Lewis et al. 2002). The fitting is improved only a little even if we add a power-law component of $\Gamma =1.65$ and $N_{\rm H}=1.5\times 10^{21}$ (Fig. 9b, reduced $\chi ^2 = 1.08$). Fig. 10 shows the energy spectrum of the central region out of the cD galaxy ($0.'5\le r \le 1.'5$). We can see a relatively high K$\beta$/K$\alpha$ intensity ratio, which is considered to be due to significant resonance scatterings. \par
Fig. 11 is the radial profile of iron abundance. The iron abundance in the central region (core of the central cD galaxy) exceeds the solar abundance and rapidly decreases with $r$. Since the value at outer region is quite normal (similar to other clusters of galaxies), the very high iron abundance should be attributed to the central cD galaxy. \par

\bigskip

\noindent{\bf 4. Conclusions}\\

\smallskip

1. There is no evidence indicating the existence of an AGN in the central region of the cD galaxy of A2029, although some hot (high hardness ratio) spots are seen within or around the cD galaxy. Our restoration of ASCA images using the Richardson-Lucy method is considered to be incomplete and gathered too many hard X-ray photons were gathered at the centeral region of the restored image.\par
2. The map of the hardness ratio over the whole cluster is not smooth but very patchy at $1.'5\le r \le 2.'5$. This region would be a transition zone between the outer region dominated by matter infalling from the outer space and the inner region in thermal equilibrium, and the patchy high-hardnass-ratio fragments would be in some violent phenomena caused by the infalling matter. \par
3. The iron abundance in the central region (core of cD galaxy) exceeds the solar value. This is consistent with the Chandra result (Lewis et al. 2002).  \par
4. The hardness ratio becomes maximum at $r\sim 2'$, but the radial temperature profile does not necessarily coincide with this. That is, the gas temperature obtained from the spectral fitting for hard (2--10 keV) photons only continues to increase in the outer regions. In addition, the degree of spectral fitting becomes worse at outer regions, when we try to fit the spectrum of 0.3--10 keV band by the thermal bremsstrahlung only. These facts shows that some additional component(s) should be added to the thermal bremsstrahlung in order to get good spectral fitting at outer regions. \par
5. The energy spectrum shows a relatively high K$\beta$/K$\alpha$ intensity ratio in the inner region but out of the central cD galaxy ($0.'5\le r \le 1.'5$). It is considered to be due to significant resonance scatterings.\par

\bigskip

\noindent{\bf References}\\

\smallskip

\noindent Allen, S.W., Fabian, A.C. The spatial distributions of cooling gas and intrinsic X-ray-absorbing material \par
in cooling flows. Mon. Not. Roy. Astron. Soc. 286, 583-603, 1997. \par
\noindent Clarke, T.E., Uson, J.M., Sarazin, C.L., Blanton, E.L. Soft X-ray absorption due to a foreground edge-on \par
spiral galaxy toward the core of A2029. Astrophys. J. 601, 798-804, 2004. \par
\noindent Dressler, A. The dynamics and structure of the cD galaxy in Abell 2029. Astrophys. J., 231, 659-670, \par
1979. \par
\noindent Grego, L., Carlstrom, J.E., Reese, E.D., Holder, G.P., Holzapfel, W.L., Joy, M.K., Mohr, J.J., Patel, S. \par
Galaxy cluster gas mass fractions from Sunyaev-Zeldovich effect measurements: constraints on $\Omega_{\rm M}$. \par
Astrophys. J., 552, 2-14, 2001. \par
\noindent Lewis, A.D., Stocke, J.T., Buote, D.A. Chandra observations of Abell 2029: no cooling flow and a steep \par
abundance gradient, ApJ, 573, L13-L17, 2002. \par
\noindent Sarazin, C.L., Wise, M.W., Markevitch, M.L. X-ray spectral properties of the cluster Abell 2029. Astro- \par
phys. J. 498, 606-618, 1998. \par
\noindent Tanioka, E., Miyoshi, S. J. Structure and X-ray spectral properties of the cluster A2029, in: Inoue, H., \par Kunieda, H. (Eds.), New Century of X-ray Astronomy. ASP Conference Series, Vol. 251, pp. 470-471, \par
2001. \par

\newpage

\noindent{\bf Figures}\\

\begin{figure}[htbp]
 \begin{tabular}{cc}
  \begin{minipage}{0.5\hsize}
  \begin{center}
    \includegraphics[width=76mm]{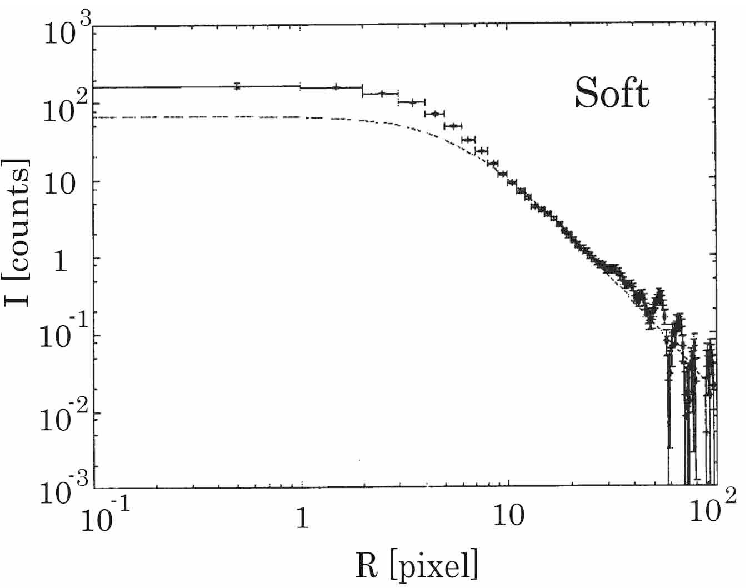}
  \end{center}
  \end{minipage}

  \begin{minipage}{0.5\hsize}
  \begin{center}
    \includegraphics[width=76mm]{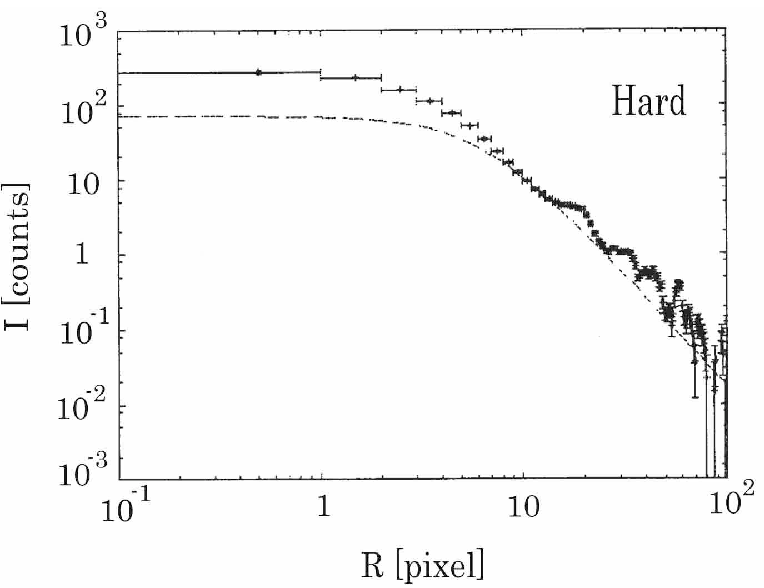}
  \end{center}
  \end{minipage}
  \end{tabular}
  \caption{$left$ Fig. 1a: Restored radial profile of the surface brightness of soft (0.7--1.5 keV) X-rays including the background (ASCA data). The dashed curve shows the best-fit isothermal $\beta $ model (1 pixel $=$ 15 arcsec). $right$ Fig. 1b: Restored radial profile of the surface brightness of hard (3--10 keV) X-rays including the background (ASCA data).  The dashed curve shows the isothermal $\beta $ model with the same values of $\beta $ and core radius as in Fig.1a (1 pixel $=$ 15 arcsec).  }
\end{figure}

\begin{figure}[htbp]
  \begin{center}
    \includegraphics[width=90mm]{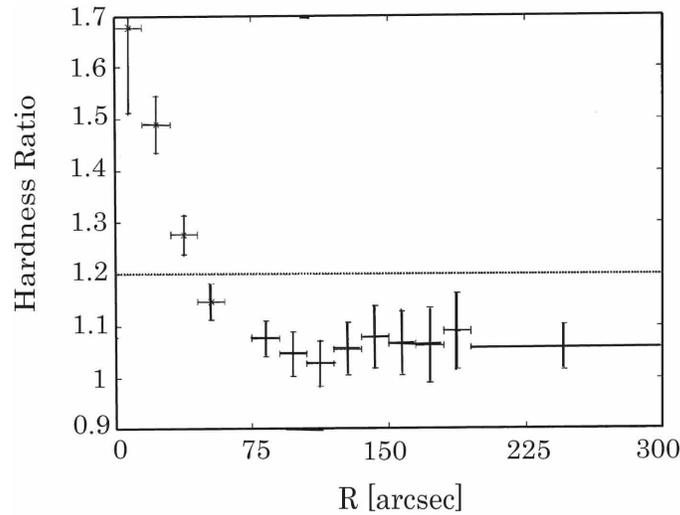}
  \end{center}
  \caption{Radial profile of the hardness ratio (3--10 keV/0.7--1.5 keV) obtained from restored ASCA data. The dashed line shows the level corresponding to $kT = 10$ keV.}
\end{figure}

\begin{figure}[htbp]
  \begin{center}
    \includegraphics[width=90mm]{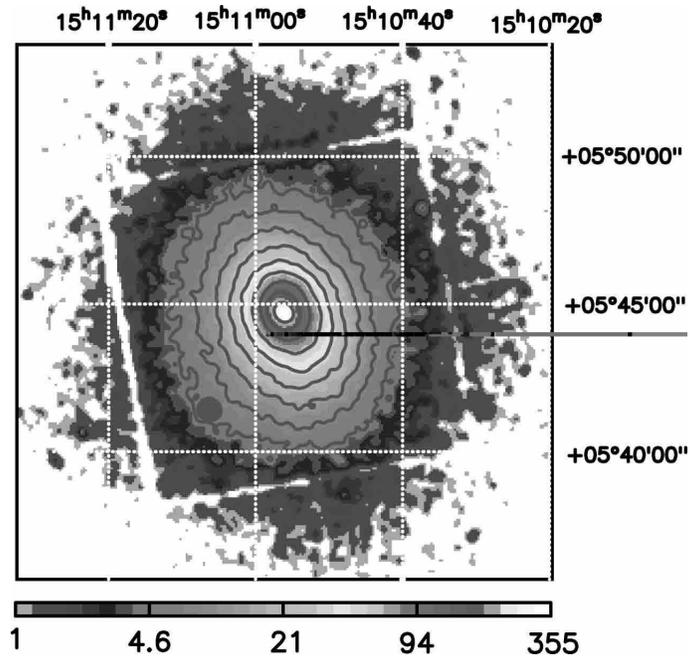}
  \end{center}
  \caption{XMM-Newton X-ray image of A2029 overlaid by the contour map (logarithmic scale; MOS1 + MOS2; 1 pixel $=$ 4.4 arcsec). Contour levels in the linear scale are 4.5, 7.5, 12.5, 20.7, 34.3, 56.9, 94.3, 156 and 259.}
\end{figure}

\begin{figure}[htbp]
  \begin{center}
    \includegraphics[width=90mm]{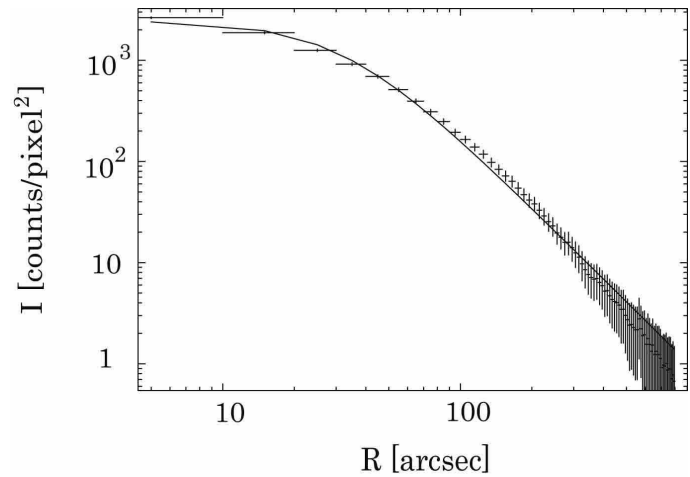}
  \end{center}
  \caption{ Radial profile of the X-ray surface brightness observed with XMM-Newton (MOS1 + MOS2, crosses) fitted by the isothermal $\beta$ model (solid line).}
\end{figure}

\begin{figure}[htbp]
  \begin{center}
    \includegraphics[width=90mm]{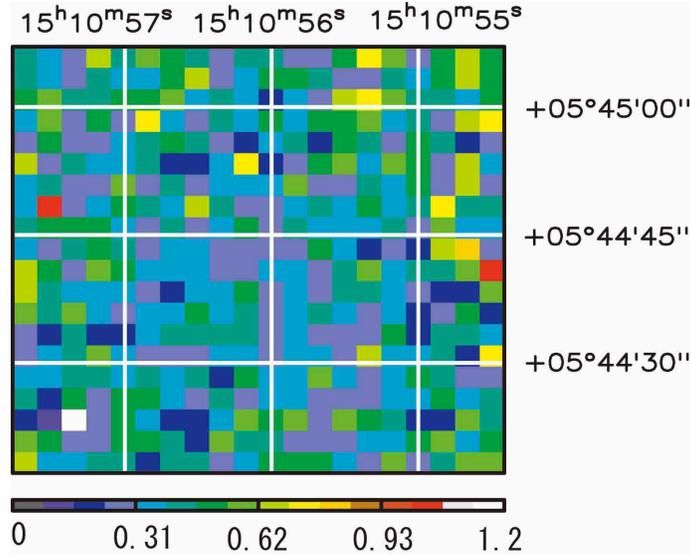}
  \end{center}
  \caption{Map of the hardness ratio (3--10 keV/0.7--1.5 keV) for the very central region ($r<17''$) of A2029 (XMM-Newton data; MOS1 + MOS2, 1 pixel $=$ 2.5 arcsec). Each of the values exceeds 3$\sigma$.}
\end{figure}

\begin{figure}[htbp]
  \begin{center}
    \includegraphics[width=90mm]{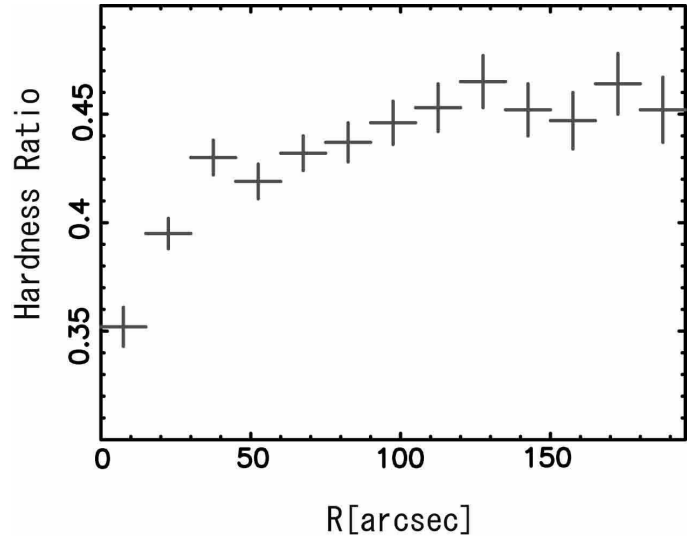}
  \end{center}
  \caption{Radial profile of the hardness ratio for XMM-Newton MOS1 + MOS2 data using the same energy bands as for ASCA data.}
\end{figure}

\begin{figure}[htbp]
  \begin{center}
    \includegraphics[width=90mm]{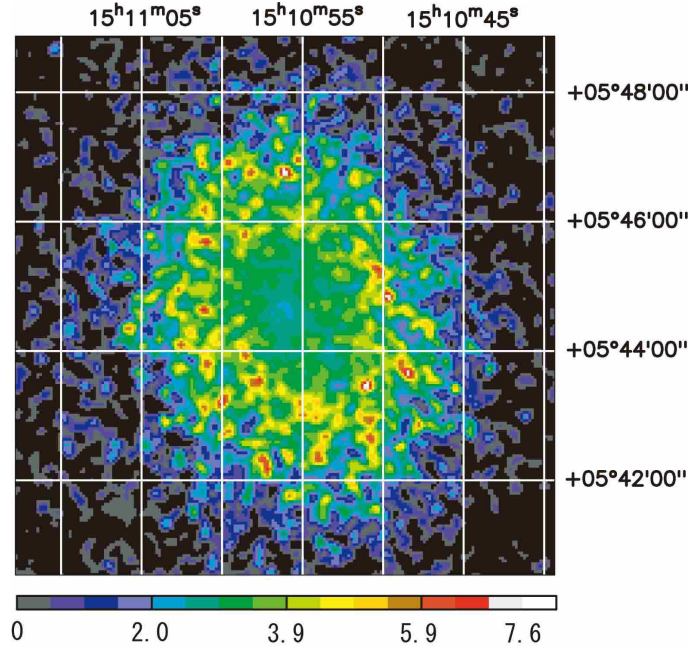}
  \end{center}
  \caption{Map of the hardness ratio over the whole cluster with vignetting correction (XMM-Newton data; MOS1 + MOS2, 1 pixel $=$ 2.5 arcsec). }
\end{figure}

\begin{figure}[htbp]
  \begin{center}
    \includegraphics[width=90mm]{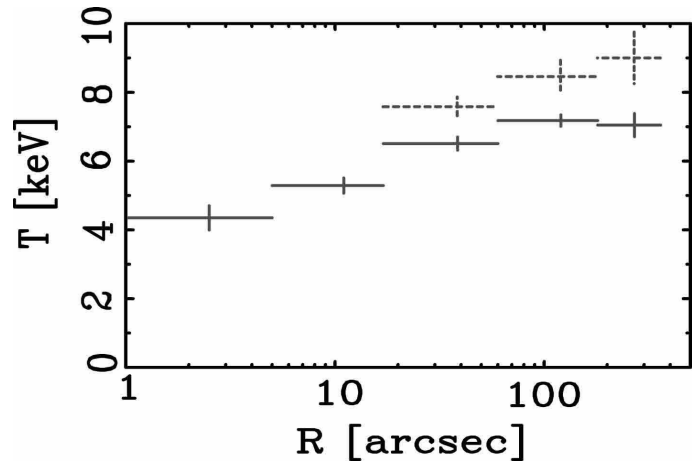}
  \end{center}
  \caption{Radial surface temperature profile obtained from the analysis of energy spectrum (XMM-Newton data; MOS1 + MOS2). Crosses indicate the gas temperatures obtained from the fitting over the whole energy range (0.3--10 keV), and dashed crosses are those for high energy (2--10 keV) photons only.}
\end{figure}

\begin{figure}[htbp]
 \begin{tabular}{cc}
  \begin{minipage}{0.5\hsize}
  \begin{center}
    \includegraphics[width=76mm]{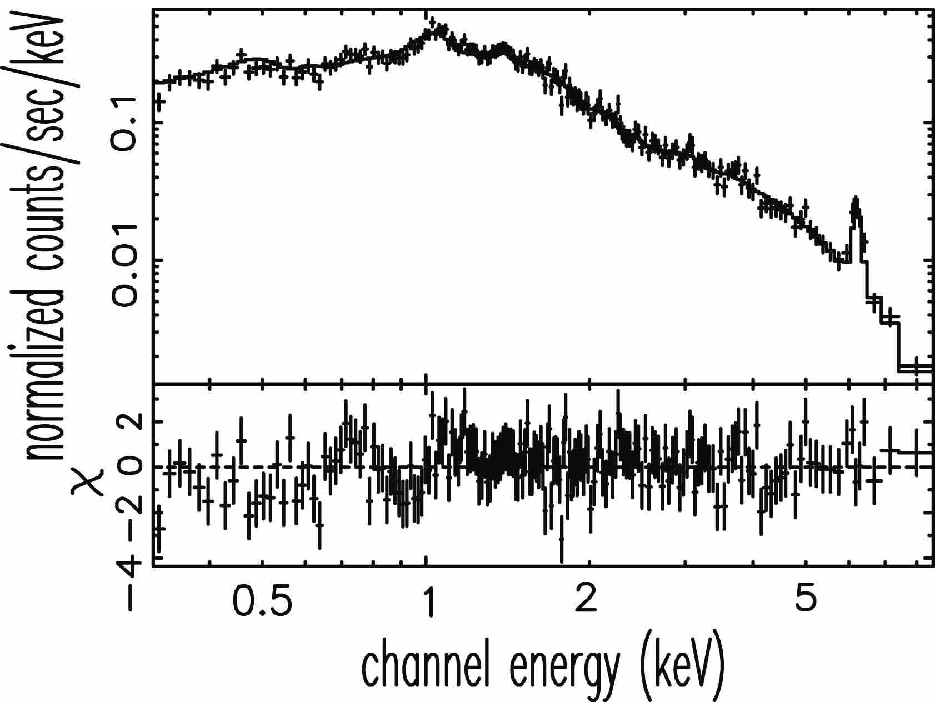}
  \end{center}
  \end{minipage}

  \begin{minipage}{0.5\hsize}
  \begin{center}
    \includegraphics[width=76mm]{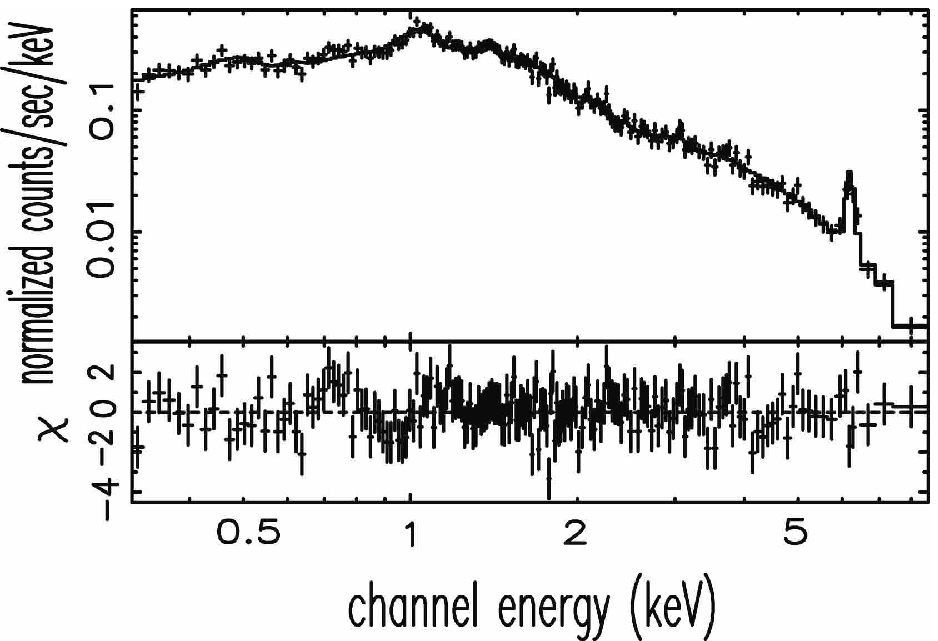}
  \end{center}
  \end{minipage}
  \end{tabular}
  \caption{$left$ Fig. 9a: The XMM-Newton X-ray energy spectrum for the inner region of $r\le 17'$ fitted by the sum of three bremsstrahlungs components of Lewis et al.'s (2002) 3D temperatures of 3.1 keV, 6.1 keV and 8.2 keV (MOS1 + MOS2, $\chi ^2$/dof $=$ 204.8/185). $right$ Fig. 9b.: The same as Fig. 9a except for adding a power law component (MOS1 + MOS2, $\chi ^2$/dof $=$ 194.6/185). }
\end{figure}

\begin{figure}[htbp]
  \begin{center}
    \includegraphics[width=80mm]{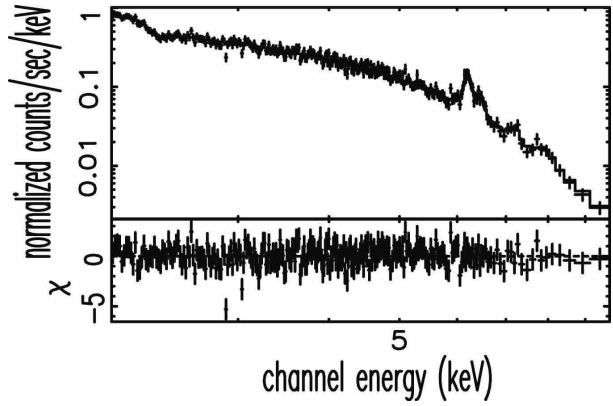}
  \end{center}
  \caption{Energy spectrum of the central region out of the cD galaxy ($0.5' \le r \le 1.5'$) (XMM-Newton data; MOS1 + MOS2, $\chi ^2$/dof $=$ 254.8/269)}
\end{figure}

\begin{figure}[htbp]
  \begin{center}
    \includegraphics[width=80mm]{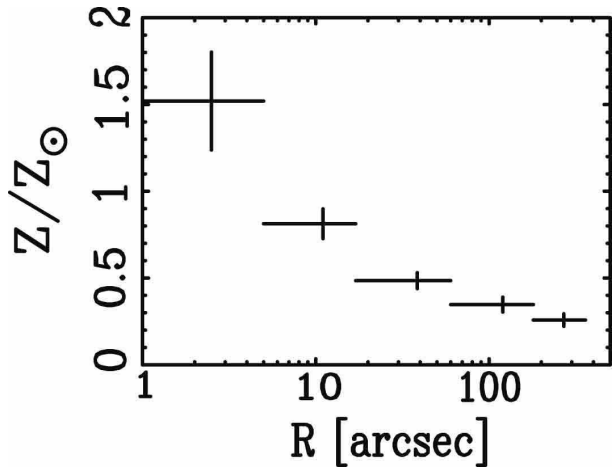}
  \end{center}
  \caption{Radial profile of the iron abundance (XMM-Newton data; MOS1 + MOS2)}
\end{figure}

\end{document}